\documentclass[prb,twocolumn]{revtex4-1}

\usepackage{amsfonts}
\usepackage{amsmath}
\usepackage{graphicx}
\usepackage{amssymb}
\usepackage[utf8]{inputenc}
\usepackage{hyperref}

\begin{document}

\title{Predicted electric field induced surface magnetization of vortex phase of superconductors}
\author{T.~M.~Mishonov}
\email{mishonov@bgphysics.eu}
\affiliation{Institute of Solid State Physics, Bulgarian Academy of Sciences
72 Tzarigradsko Chaussee Blvd., BG-1784 Sofia, Bulgaria}
\affiliation{St. Clement of Ohrid University at Sofia, 5 James Bourchier Blvd., BG-1164 Sofia, Bulgaria}

\author{V.~I.~Danchev}
\affiliation{St. Clement of Ohrid University at Sofia, 5 James Bourchier Blvd., BG-1164 Sofia, Bulgaria}

\author{A.~M.~Varonov}
\email{varonov@issp.bas.bg}
\affiliation{Institute of Solid State Physics, Bulgarian Academy of Sciences
72 Tzarigradsko Chaussee Blvd., BG-1784 Sofia, Bulgaria}

\date{17 February, 2020, 16:30}

\begin{abstract}
We predict a new effect in condensed matter surface magnetization of the vortex phase of a superconductor induced by electric field.
The magnetized superconductor should be one of the plates of a plane capacitor on which a voltage has to be applied. 
Applying alternating voltage to the capacitor, electrostatic induction leads to oscillations of the magnetic moment which has to be measured by electromotive force inducted in a detector coil.
The derived explicit formula for the magnetization contains the effective mass of Cooper pairs 
and a systematic investigation of the predicted magnetization will lead to a creation of an effective 
Cooper pair mass spectroscopy.
For cleaved superconductors this effective mass is a property of the bulk material.
\end{abstract}
\maketitle

Can an AC dielectric induction create AC magnetization 
on the surface of an superconductor in the vortex
Schubnikov-Abrikosov phase?
The purpose of the present work is to predict a new effect in the physics of supreconductivity which can be easily observed using standard electronic equipment -- excess magnetic moment induced by the excess charge on the surface of magnetized type-II superconductor.

Let us consider a superconducting film or slab of 
type-II superconductor, for example YBa$_2$Cu$_3$O$_{7-\delta}$ or 
Bi$_2$Sr$_2$CaCu$_2$O$_8$ with thickness $d_{\mathrm{film}}$ and surface area $S$ in an external electromagnetic field and a superconducting current around 
a single Abrikosov vortex formed within the film.
Our starting point is the well-known Bohr quantization of angular momentum 
which describe the velocity distribution of the superfluid\cite{Onsager}
\begin{eqnarray}
\label{Sommerfeld}
m^{*} r v = \hbar,
\end{eqnarray}
where $m^{*}$ is the effective Cooper pair mass in the plane of the film, $r$ is the radial distance from its center and $v$ is the velocity of circulating fluid of Cooper pairs.
For a nice introduction of basic physics of superfluids we recommend the books 
by Abrikosov\cite{Abrikosov} and Feynman.\cite{Feynman}
Equation~(\ref{Sommerfeld}) is valid in the range $\xi \ll r \ll \lambda$, where $\xi$ is the coherence length and $\lambda(T)$ is the temperature dependent penetration depth\cite{LL9,deGennes,London,GL}
\begin{eqnarray} \label{penetration}
\frac{1}{\lambda^2(T)}=\frac{e^{*2}n_{_{3D}}(T)}{\varepsilon_0 c^2 m^*},
\end{eqnarray}
where in Gaussian system $\varepsilon_0=1/4\pi.$
Let us mention 
vortices are property of the magnetized phase of type-II superconductor below $T_c$.
We suppose also that $\varkappa=\lambda/\xi\gg1.$
The magnetic moment per unit film thickness of this single vortex is
\begin{eqnarray}\label{moment1}
\frac{M_1}{d_{\mathrm{film}}}=\int_{\xi}^{\lambda} \pi r^2  j\mathrm{d} r
=\int_{\xi}^{\lambda} \pi r^2  n_{\mathrm{_{3D}}} e^{*} v\mathrm{d} r,
\end{eqnarray}
where $j=n_{\mathrm{_{3D}}} e^{*} v$, cf. Ref.~\onlinecite{Abrikosov}.
We consider thick films $d_{\mathrm{film}}>\lambda$ or even slabs of type-II superconductors.
Expressing $v(r)$ from the Bohr quantization condition Eq.~(\ref{Sommerfeld}) and substituting it in  Eq.~(\ref{moment1}), after integration we obtain the magnetization of a single vortex per unit film thickness
\begin{eqnarray}\label{M1}
\frac{M_1}{d_{\mathrm{film}}}
=\frac{n_{\mathrm{_{3D}}}\pi \hbar}{2}\left(\frac{e^{*}}{m^{*}} \right) (\lambda^2 - \xi^2).
\end{eqnarray}
The magnetic moment of a single vortex piercing the superconducting slab is $M_1$ and 
$M_1/d_{\mathrm{film}}$ has dimension magnetic moment per unit length.
We can neglect $\xi^2$ in Eq.~(\ref{M1}) as $\xi$ is assumed to be a small quantity,
i.e. we suppose type-II superconductor with $\varkappa\equiv\lambda/\xi\gg1.$
For magnetic fields much smaller than the upper critical field $B<B_{c2}$
the vortices with area density $B/\Phi_0$ and total number $N=BS/\Phi_0$ are weakly overlapping and the total magnetic moment of the sample is given by the sum of individual magnetic moments
\begin{eqnarray}
\mathcal{M}=N M_1 =\frac{\Phi}{\Phi_0} M_1 = \left(\frac{M_1}{d_{\mathrm{film}}} \right) \frac{B}{\Phi_0} V,
\end{eqnarray}
where we recognize the volume of the film $V=d_{\mathrm{film}} S$ 
and the total magnetic flux of the magnetic field (assumed orthogonal to the film) $\Phi=BS$.
For the magnetic moment per unit area of the sample we have
\begin{eqnarray}\label{specificM}
\frac{\mathcal{M}}{S}=M_1 \frac{B}{\Phi_0}
=\frac14 \left(\frac{e^{*2}}{m^{*}}\right) n_{\mathrm{_{3D}}}  d_{\mathrm{film}} \lambda^2(T) B,
\end{eqnarray}
where $\Phi_0=2 \pi \hbar/{|e^*|}$ is the flux quantum 
and $e^*$ is the Cooper pair charge 
$|e^{*}|=2|e|$.
One additional speed of light $c$ has to be added in the denominator in Gaussian system.
The multiplier $n_{\mathrm{_{2D}}}(T) = d_{\mathrm{film}} n_{\mathrm{_{3D}}}(T)$
is the temperature dependent two dimensional (2D) density of the superfluid charge carriers.
As pointed out above, this result for the magnetization is a good approximation when the magnetic field is much smaller than the upper critical field $B\ll B_{c2}(T).$
Substituting Eq.~(\ref{penetration}) in the approximation Eq.~(\ref{specificM}) and recognizing once again the volume of the film, we arrive at the simple result for the magnetization 
\begin{eqnarray}
M\equiv\frac{\mathcal{M}}{V}\approx \frac{B}{4\mu_0},
\qquad \mu_0=1/\epsilon_0 c^2.
\end{eqnarray}
This law can be observed in the field cooling regime.
For comparison for Meissner-Ochsenfeld (MO) phase we have
$M_\mathrm{MO}=-B/\mu_0.$
In Gaussian system we have to substitute $\mu_0=4\pi$
while in Lorentz–Heaviside units $\mu_0=1$ and $\epsilon_0=1$.
For a slab in perpendicular field the demagnetizing factor is $n=1$
and for all magnetic fields $B$ in interval $(0,\;B_{c2})$ the sample is in vortex state.

\begin{figure}[h]
\includegraphics[scale=0.4]{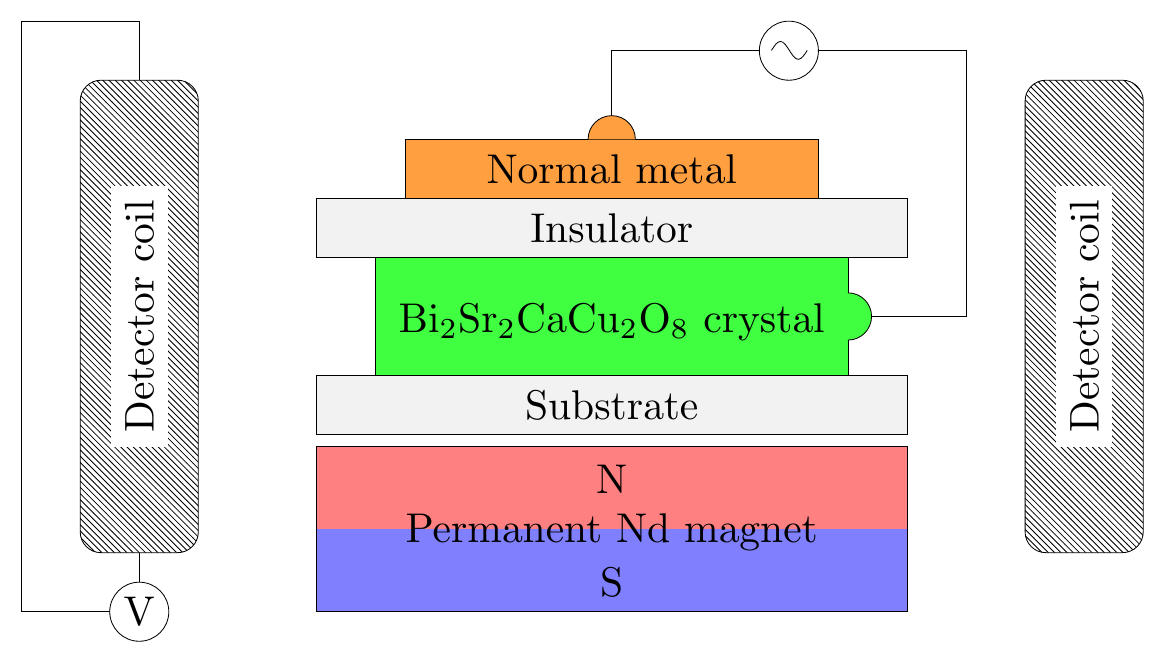}
\caption{
A schematic representation of the proposed setup.
A crystal slab of Bi$_2$Sr$_2$CaCu$_2$O$_8$ is connected with an alternating voltage source through a lateral contact.
The cleaved surface of the superconductor forms one of the plates of a capacitor.
The superconductor surface is covered by an insulator, 
on which a metal layer is evaporated, acting as the second plate of the capacitor. 
The system is placed in an external magnetic field, which is perpendicular to the capacitor plates.
The magnetic moment is detected by a detector coil of a magnetometer.
The U-shaped ferrites, signal generator, preamplifier and lock-in are not depicted.}
\label{Fig1}
\end{figure}

The set-up we propose consists of a plane capacitor whose plates are perpendicular to an external magnetic field, shown in Fig.~\ref{Fig1}.
The capacitor is placed in a detector coil.
One of the capacitor plates is the superconductor with an atomically clean surface, while
the other plate is unspecific normal metal.
An AC voltage is applied to the plates, resulting in a modulation of the Abrikosov lattice magnetic moment, which can be registered by the detector coil.

The scenario considered has one crucial difference as compared to that leading to Eq.~(\ref{specificM}),  the voltage difference makes the whole film act as a capacitor and there is additional 2D charge density being accumulated due to this field.
Recalling the connection between surface charge density $\rho_{\mathrm{_{2D}}}$ and the normal component of electric induction $\mathbf{D} \cdot \mathrm{d} \mathbf{S}=\rho_{\mathrm{_{2D}}} |\mathrm{d}\mathbf{S}|$ and simplifying it in the case of our geometry, we see that the additional charge accumulated will be the $z$-component of the electric induction $D_z$ for a temperature tending to 0 (we have assumed the $z$-component to be parallel to the film thickness).
In the case of a finite temperature, we can form the dimensionless ratio, which coincides with the percent of the superfluid carriers
\begin{equation}
C(T)=\frac{n_{\mathrm{_{2D}}}(T)}{n_{\mathrm{_{2D}}}(0)}=\frac{\lambda^2(0)}{\lambda^2(T)}.
\end{equation}
This ratio determined by the penetration depth is a tool for interpretation of the experimental data
known as two fluid model. 
Recalling once again the formula for the penetration depth given by Eq.~(\ref{penetration}),
Eq.~(\ref{specificM}) should be corrected simply by a substitution of the two dimensional (2D) superfluid charge per unit area, which is the modulated electric field 
\begin{eqnarray}
e^*n_{\mathrm{_{2D}}}(T) = e^* d_{\mathrm{film}} n_{\mathrm{_{3D}}}(T)+D_z \frac{\lambda^2(0)}{\lambda^2(T)}.
\end{eqnarray}
Here we can recognize the total electric charge induced on the superconductor surface
$Q=SD_z$.
The microscopic approach gives different functions for 
the temperature dependent superfluid ratio $C(T)$
but the final formulas for $m^*$ are invariant.
The electric charges accumulated on the superconducting surface are partially superfluid 
and contribute to the magnetic field of the vortices;
$C(T)D_z$ is the superfluid surface charge density.
Applying alternating current to the superconductor surface, we have
$I(t)=\omega Q_0\cos (\omega t)$ modulating charge oscillations
$Q(t)=Q_0 \sin (\omega t)$ and simultaneously measure the time dependent magnetic moment 
$M(t)=M_0 \sin (\omega t)$.
The effective mass of Cooper pairs can be expressed by the ratio of the amplitudes
\begin{equation}
m^{*}=\dfrac{e^* B \lambda^2(0)}{4  \dfrac{\mathrm{d} M}{\mathrm{d} Q}} 
=\frac{e B \lambda^2(0) Q_0}{2\,\mathrm{sgn}(e^*) M_0},
\label{effective_mass}
\end{equation}
where the derivative here is at constant magnetic field and the ratio $M_0/Q_0$ is actually the slope of a linear regression when the electrostatic doping is due to different voltages.
The cancellation of the temperature dependent $\lambda^2(T)$ suggests that the parameter determined in this way will have weak temperature dependence.
This result allows us to measure the effective Cooper pair mass $m^{*}$ through the experimental setup presented in Fig.~\ref{Fig1} by measuring the change in the magnetic moment of the sample induced by the accumulation of charges due to a voltage generator, all of which is happening in the external magnetic field $B$ perpendicular to the film. 
Since the effective mass is positive, 
the method simultaneously determines the sign of the Cooper pair charge 
$\mathrm{sgn}(e^*)$. 
If an investigated superconductor is used for which the penetration depth
$\lambda(0)$ is known, all variables from the right side of Eq.~(\ref{effective_mass}) are known
and a systematic investigation of the predicted surface magnetization
can lead to a creation of Cooper pair mass spectroscopy; 
which is one of the purposes of the present work. 
For cleaved superconductors $m^*$ coincides 
with the bulk one while 
in general case $m^*$ will reveal some properties 
of the superconductor-insulator interface.


For which effects of the physics of superconductivity effective mass of Cooper pairs is important?
Except the effect of Bernoulli in superconductors, which is only qualitatively observed,
$m^*$ is important for electric modulation of the kinetic inductance in thin high-$T_c$
films. 
In any cases the relative mass $m^*/2m_e$
will be a convenient dimensionless parameter 
parameterizing the new effect.

First quantitative determination of effective mass of Cooper pairs 
in high-$T_c$ cuprate YBa$_2$Cu$_3$O$_{7-\delta}$ 
in vortex-free regime
was performed in Bell labs 32 years ago,~\cite{Reply} cf. Ref.~\onlinecite{Comment}.


From a technological point of view, we emphasize that it is not necessary to make contacts on the ab-plane of Bi$_2$Sr$_2$CaCu$_2$O$_8$.
This face, which is easily cleaved, is one of the plates of the already mentioned plane capacitor and it has to be atomically clean.
The contact has to be made on the lateral ac- or bc-plane, which is a relatively simple task.
The dielectric material between the plates of the capacitor does not have any specific contribution to the effect, which should be observable even for plain air.
In such a way, the proposed experiment can be easily performed in every laboratory related to the physics of superconductivity, it is only necessary to have a single Bi$_2$Sr$_2$CaCu$_2$O$_8$ crystal.

Let us give a numerical example.
First of all let us try to minimize the parasitic cross-talk between the driver and detector coil using a small tun-able 
mutual inductance. 
At cooling below $T_c$ the searched effect  
will appear as smeared jump of the amplitude of magnetization.
If a variable magnetic moment  $\mathcal{M}(t)$ is placed in a solenoid with $\nu$ turns per unit length the induced electromotive force in the solenoid is 
$\mathcal{E}_\mathrm{coil}=-\mu_0 \mathrm{d}\mathcal{M}(t)/\mathrm{d}t.$
Using a 250~$\mu\mathrm{m}$ copper wire it is possible to make a coil with effective 
$\nu=1.37\times 10^5$~turns/m,
Let us use $S=4\,\mathrm{mm}^2$ sample of Bi$_2$Sr$_2$CaCu$_2$O$_8$. 
If at the cleaved surface we apply polyethylene dielectric relative
susceptibility $\varepsilon_\mathrm{rel}=2$ with breakdown voltage $E_\mathrm{bd}= 10$~MV/m,
we can apply $E=10\,\mathrm{MV/m}$, 
the electric induction is $D=\varepsilon_0\varepsilon_\mathrm{rel}E$ and the amplitude of the charge
is $Q_0=DS=708\,\mathrm{pC}$.
For $d_\mathrm{ins}=10~\mu\mathrm{m}$ polyethylene film the amplitude of the charging voltage is
$U_\mathrm{drive}=100\,\mathrm{V}$ and the capacity is 
$C_\mathrm{cap}=\varepsilon_0\varepsilon_\mathrm{rel}S/d_\mathrm{ins}=7.08~$pF.
At $f$=50~kHz frequency and $\omega=2\pi f$ the amplitude of the charging current is 
$I_0=\omega Q_0=222\,\mu\mathrm{A}$.
For the source of external magnetic field we can use a permanent Nd 
magnet with magnetization $B=1\,\mathrm{T}$.
For evaluation we can use for the effective mass of Cooper pairs $m^*=10 \,m_e$, where $m_e$ is the electron mass.
For these parameters according to Eq.~(\ref{effective_mass}) the amplitude of the magnetization is
$\mathcal{M}_0=62~\mathrm{fA\,m^2}.$
This AC magnetization is induced in the detector coil with a voltage with amplitude
$U_0=\mu_0 \nu \omega M_0=3.36$~nV.
This signal has to be amplified million times $Y=10^6$
by a low noise pre-amplifier and additionally 10 times by a resonant amplifier.
Then the amplitude of the amplified signal can reach $U_\mathrm{ampl}=Y U_0= 33.6$~mV, which can be measured by a lock-in voltmeter.
This magnetic moment is however only 14\% from the largest parasite magnetic moment related to the charging of the capacitor $\mathcal{M}_\mathrm{par}\cong S I_0/100$, where the denominator 100 is order evaluation.
The parasitic mutual inductance will be minimized if the capacitor is charged by a coaxial cable.
Additionally using variable mutual inductance this cross-talk has to be carefully annulled above $T_c.$
Additionally, those two magnetizations are phase shifted at 90$^\circ$.
The magnetization of the new effect is proportional to the charge and voltage, while the parasite magnetic moment is proportional to the current, i.e. charge derivative.
The true effect can be easily distinguished as a constant term making linear regression at different frequencies in the plot $\mathcal{E}_\mathrm{coil}^2$ versus $\omega^2$ if the experiments are done at fixed $I_0.$
The coils with large $\nu=137$~turns/mm and linear size of several cm have also a large Ohmic resistance, in our example $R_\mathrm{coil}\approx 402~\Omega$.
The magnitude of the thermal noise can be evaluated as
$U_\mathrm{noise}= \sqrt{4 k_\mathrm{B} T \Delta f}$,
where $k_\mathrm{B}$ is the Boltzmann constant, $T=80$~K at boiling nitrogen and 
$\Delta f$ is the frequency bandwidth.
If at low frequency the signal is collected for 20 minutes, we can take as order evaluation 
$\Delta f\simeq 1\,\mathrm{mHz}.$
Only for such long times of signal evaluation we can reach signal to noise ratio
$U_0/U_\mathrm{noise}\approx 4$.
Additionally the detector coil has to be carefully screened from the external electromagnetic noise since it is a perfect antenna.
In short, the experiment should be carefully performed but it is doable.

The effective mass of a charge carrier is a parameter of a dynamic theory and requires electric field due to the gauge-invariant time derivatives $\mathrm{i} \hbar \partial_t - e \varphi$.
The electric field should be present on the interface between a superconductor and an insulator, and the proposed experiment requires extremely clean interfaces.

The proposed experiment utilizes the electrostatic doping of fluxons surface to measure this crucial parameter in the physics of superconductivity.
A carefully prepared structure of Kapton on YBa$_2$Cu$_3$O$_{7-\delta}$ was the system used for the first determination of the effective mass of Cooper pairs. \cite{Reply}
We believe, however, that Bi$_2$Sr$_2$CaCu$_2$O$_8$, which is easily cleaved by adhesive tape surfaces, will catalyze the development of Cooper pair mass spectroscopy.
Here we have to point out that the electron band structure of over-doped high-$T_c$ cuprates 
is easily fitted by Linear Combination of Atomic Orbital (LCAO) 
method.~\cite{Mishonov:00,Mishonov:03,Stoev:10}
The discrepancy between the corresponding integral for optical mass that can be easily calculated and the experimentally determined Cooper pair mass will reveal how good we understand the effects of strong correlations.
We suppose for the pnictide superconductors the situation is the same. 

The next step would be a systematic investigation of thick films $d_{\mathrm{film}} > \lambda$ of 
technological superconductor alloys.
Historically, the effective mass of Cooper pairs is implicitly introduced in the seminal papers by the London brothers~\cite{London} and Ginzburg-Landau (GL).~\cite{G-L}
Further on, Gor'kov derived GL theory from microscopic approach.
Later Gor'kov has extended his treatment to include finite mean-free-path effects.
He has found that the equations have the same form as above, except that the 
``mass'' $m^*$ is increased to that of pure metal.~\cite{Schrieffer}
For conventional superconductors with strong disorder,
the effective mass is proportional to the Ohmic resistivity $m^*\propto\varrho_{_\Omega}$.
As the Pippard kernel well approximates the BCS one, this proportionality law
can be understood in the framework of Pippard-Landau theory.~\cite{Mishonov:94a}
For strongly disordered superconducting films, for which the temperature of 
Kosterlitz-Thouless transition $T_\mathrm{KT}$ is significantly below the BCS one $T_c$, 
the effective mass $m^*$ reaches hadronic values.
For layered high-$T_c$ cuprates with strong anisotropy giant effective mass in c-direction
$m^*_c$ was a tool to predict a superconducting collective mode plasma oscillations in the
Bi$_2$Sr$_2$CaCu$_2$O$_8$ high-temperature superconductor.~\cite{Mishonov:91}
Superconducting collective modes predicted for thin superconducting films~\cite{Mishonov:90}
give another way to determine Cooper pair mass;~\cite{Mishonov:94a} 
an alternative is to use Bernoulli effect in superconductors.~\cite{Mishonov:94b}

In conclusion, using magnetic fields only, it is possible to determine only the ratio
$n_{_{3D}}(T)/m^*$. 
In order to determine $m^*$ separately, it is necessary essentially to use electric field.
This parameter is inaccessible by routine methods as specific heat, surface impedance, 
penetration depth etc.
In some sense in the present paper we predict a new electric effect in superconductors
-- electric field induced surface magnetization of the vortex lattice.
Our final result Eq.~(\ref{effective_mass}) can be rewritten as 
\begin{equation}
\mathrm{d} cM=\frac{e^* }{4  m^*}\lambda^2(0)B\, \mathrm{d} Q
=\frac14 \frac{e^* B}{m^*}\,\mathrm{d}Q\,\lambda^2(0).
\label{Excess_magnetic_moment}
\end{equation}
Without the dimensionless factor 1/4 final formula for the excess magnetization can be 
understood by model estimation: ``cyclotron'' frequency 
$e^* B/m^*c$ ($c$ is only in Gaussian units) times charge $\mathrm{d}Q$
gives current, 
which multiplied by area $\lambda^2(0)$ gives magnetic moment $\mathrm{d} M$.
We concluded that cleaved Bi$_2$Sr$_2$CaCu$_2$O$_8$ surface will be the best material 
to observe the new effect and then it become part of the material science of superconductors.

\textbf{Acknowledgments.} This work was partially supported by European COST activity NANOCOHYBRI-2018 and was presented in the Vortex 2019 conference on 20--25 May in Antwerp, Belgium.
Authors are grateful to Alexander~Backs, Daniele~Torsello, Leonardo~Civale, Milorad~Milo\v{s}evi\'{c}, Tsuyoshi~Tamegai, Beno\^{i}t~Vanderheyden, Alexei~Koshelev, Ritika~Panghotra and Vladimir~Krasnov for stimulating discussions and interest towards the research.
The recent stimulating correspondence with Dr.~Octan Pachov is also highly appreciated.

\end{document}